# Relativistic Effects on X-ray Structure Factors

**Kilian Batke, Georg Eickerling**

E-mail: `georg.eickerling@physik.uni-augsburg.de`
Institut für Physik, Universität Augsburg, Universitätsstraße 1, D-86159 Augsburg, Germany

**Abstract.** X-ray structure factors from four-component molecular wave functions have been calculated for the model systems $M(C_2H_2)$ ($M$= Ni, Pd, Pt). Relativistic effects on the structure factors are investigated by the comparison to results obtained from a non–relativistic reference and in order to systematically analyse the effect of different quasi-relativistic approximations, we also included the DKH2 and the ZORA Hamiltonian in our study. We show, that the overall effects of relativity on the structure factors on average amount to 0.47, 0.80 and 1.27% for the three model systems under investigation, but that for individual reflections or reflection series the effects can be orders of magnitude larger. Employing the DKH2 or ZORA Hamiltonian takes these effects into account to a large extend, reducing the according differences by one order of magnitude. In order to determine the experimental significance of the results, the magnitude of the relativistic effects on the structure factors is compared to the according changes due to charge transfer and chemical bonding.

## 1. Introduction

Molecular X-ray structure factors calculated from wave functions obtained by *ab-initio* methods play an important role when investigating the accuracy of models employed for the reconstruction of electron density distributions $\rho(\mathbf{r})$ from experimental data. The most commonly used ansatz in this respect, the *Hansen-Coppens Model* (HC) [1], has been evaluated on this basis many times (see, for example refs. [2–6]) and has more recently been generalized to an *Extended Hansen-Coppens Model* (EHC). [7,8] Employing calculated structure factors $\mathbf{F}_c(\mathbf{r}^*)$ for such studies provides reference data free of systematic experimental errors and effects such as absorption or extinction. In addition, the $\mathbf{F}_c(\mathbf{r}^*)$ are based on static electron density distributions and therefore not affected by dampening effects due to thermal motion of the atoms. This provides a reference when assessing the degree of deconvolution of thermal motion and chemical bonding effects on the electron density. Finally, varying the level of approximation employed for the quantum chemical calculations allows for a systematic study of effects such as electron correlation [9–15], basis set size [6,16] or the type of model Hamiltonian on the resulting structure factors and electron densities. [9,17–23]

The aim of the present study is a systematic investigation of relativistic effects on X-ray structure factors. While this topic has been studied for atomic form factors $f(r^*)$ in the past, [24–27] there exists up to date no thorough study in this respect on structure factors $\mathbf{F}(\mathbf{r}^*)$ based on a fully-relativistic reference. However, results from a previous study on relativistic effects on the topology of the electron density suggest, that especially for third row transition metal elements relativistic effects need to be taken into account in order to obtain accurate electron density distributions. [17] Due to the close entanglement of theory and experiment in the process of obtaining experimental charge density distributions $\rho(\mathbf{r})$ from measured structure factors $\mathbf{F}_o(\mathbf{r}^*)$ [28] one aim of the present study is to assess the effects of employing different (quasi–)relativistic model Hamiltonians for the calculation of molecular structure factors.

The investigation of relativistic effects, especially when aiming at a later interpretation with respect to experimental results, requires the definition of a suitable non–relativistic (NR) reference. Following for example the definition by Reiher and Wolf, the term *relativistic effects* can be defined as "*the difference between relativistic and non–relativistic expectation values*". [29, p. 555] As such, relativistic effects are not measurable because a non–relativistic experiment cannot be performed. Therefore, their experimental determination can only be based on the comparison of the measurement of a relativistic observable and a non–relativistic reference expectation value obtained from quantum chemical calculations. In a theoretical study, this definition imposes no additional difficulties, as relativistic effects may be discussed by comparing results from a (quasi–)relativistic to those of a non–relativistic calculation. The choice of the according model Hamiltonians determines, to which extend relativistic effects will be accounted for in a given study.

Due to the quasi closed-shell nature of the model compounds under investigation within our study (*vide infra*), we focus on the *scalar–relativistic effects*. [23] For the purpose of our discussion of X-ray structure factors, these are dominated by the relativistic contractions of the inner electronic shells of the atoms due to the non-classical speed of the according electrons. For isolated atoms, one may estimate the ratio of the non-relativistic and relativistic orbital radius ($r_0$ and $r_r$, respectively) in atomic units according to

$$\frac{r_0}{r_r} = \left[1 - \left(\frac{Z}{n \cdot c}\right)^2\right]^{-\frac{1}{2}} = \frac{m_r}{m_0} \quad (1)$$

by considering the relativistic mass increase of the electrons (see, for example [30–32]). This ratio can be expressed by the atomic number $Z$, the main quantum number $n$ of an atomic orbital and the speed of light $c$ or by the ratio of the relativistic mass $m_r$ compared to the mass $m_0$ at zero velocity. Equation (1) yields an estimated relativistic contraction $\Delta r = r_0/r_r$ of the $1s$ atomic shell by 2, 6 and 22% for a Ni, Pd and Pt atom, respectively. The according values may be verified on the basis of scalar–relativistic atomic *ab–initio* calculations by determining the shift of the outermost local maxima in the radial distribution function $D(r) = 4\pi r^2 \rho(r)$ of the individual electronic shells defined by their main quantum number $n$ (see Figure 1). From these shifts one obtains a relativistic contraction of $\Delta r =$ 21.3, 9.9, 4.6 and 2.5% for the $n =$ 1, 2, 3 and 4 shell of a Platinum atom, respectively, where the value of 21.3% is in very good agreement with the estimated value of 22% specified above. We note at this point, that within the present study we will not consider the different radial extensions of the *spin–orbit* coupled spinors. For details on the different radial behavior of the spinors, see, for example References [33–35].

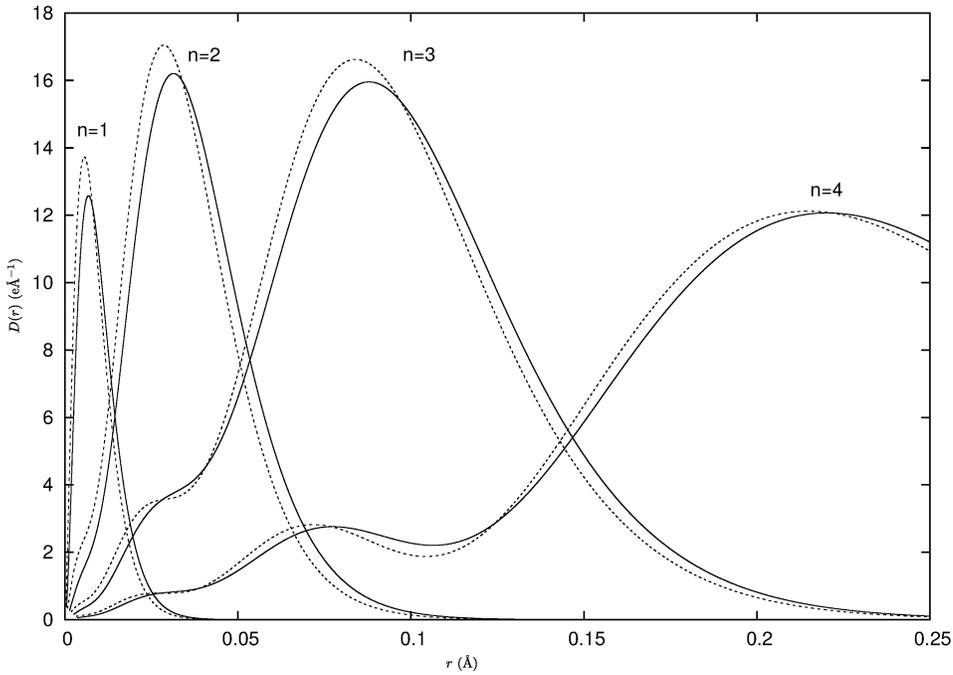

**Figure 1.** $n$-resolved radial distribution function $D(r)$ of a Platinum atom as obtained from non–relativistic HF calculations and the according calculations employing the DKH2 Hamiltonian (solid and dashed lines, respectively) *vs.* the distance from the atomic position in Å.

Further complications of the study of relativistic effects may arise due to the change of picture for approximate relativistic Hamiltonians. [35] Recently, Bučinský *et al.* studied this *picture change effect* (PCE) with respect to X-ray structure factors when employing the DKH2 Hamiltonian. [20] The results of this study on Copper complexes showed, however, that these effects are one order of magnitude smaller than the overall relativistic effects. In our study, we will therefore not explicitly consider the PCE.

One important aspect of the investigation of relativistic effects on the basis of X-ray structure factors (instead of directly on the electron density) is the intrinsic dependency of the results on the data resolution available. The data resolution of an X-ray diffraction experiment is usually given either as the shortest lattice–plane distance $d$ obtained for a given maximum Bragg diffraction angle $\theta$ and a fixed wavelength $\lambda$ from the Bragg

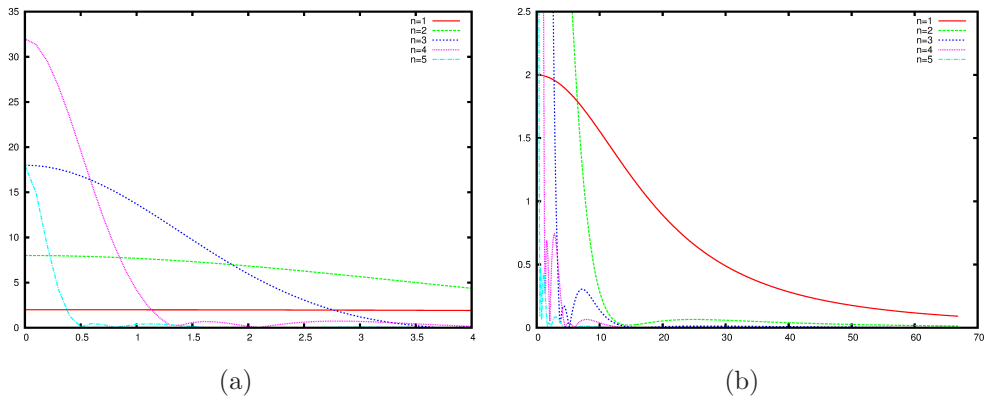

**Figure 2.** Contributions of the electronic shells $n$ to the atomic form factor $|f_{DKH2}(r^*)|$ of a Platinum atom *vs.* the resolution $\sin\theta/\lambda$ in Å$^{-1}$; (a) data resolution $\leq 4$Å$^{-1}$, (b) data resolution $\leq 67$Å$^{-1}$.

equation, [36] or (the notation used throughout the present study) in terms of

$$\frac{\sin\theta}{\lambda} = \frac{1}{2d} \tag{2}$$

in units of Å$^{-1}$. This reciprocal distance exemplifies the relation between real and reciprocal space in the scope of the diffraction experiment, a point that may be further illustrated by the according relations between the electron density distribution in real space and the structure factor in reciprocal space. In the following, we will therefore briefly summarize the most important consequences of this inverse relation with respect to the investigation of relativity for atomic form factors.

Due to the small radial extend of the innermost electronic shells in real space, we may expect relativistic effects on scattering factors to be most significant at high data resolutions, *i.e.* at high scattering angles $\theta$ or high values of $\sin\theta/\lambda$. To illustrate this fact, Figure 2 depicts the atomic scattering factor contributions of the electronic shells of an isolated Platinum atom as defined by the quantum number $n$ *vs.* the data resolution in Å$^{-1}$. The graph limited to $\sin\theta/\lambda \leq 4$Å$^{-1}$ (Figure 2a) clearly points out the inverse relation between the electron density distribution (and also of $D(r)$, see Figure 1) and the atomic scattering factor by the almost constant, *i.e.* extremely diffuse contribution of the $1s$ shell in reciprocal space and thus extremely contracted inner most shell in real space. Note, that in order to resolve the scattering angle dependency of the $1s$ shell, data resolutions up to $\sin\theta/\lambda \approx 70$Å$^{-1}$ are required (Figure 2b). We may for example conclude from this, that relativistic effects in the $1s$ shell of a Platinum atom will not be accessible by standard X-ray diffraction experiments nowadays, as their maximum resolution is limited to approx. $\sin\theta/\lambda=4$Å$^{-1}$ even at third-generation synchrotrons. [37]

In order to assess the relativistic effects on the individual atomic shells as revealed by the atomic scattering factor, Figure 3a depicts the difference $|f_{DKH2}(r^*)|-|f_{NR}(r^*)|$ for a Platinum atom for $\sin\theta/\lambda \leq 67$Å$^{-1}$. Contrary to the statement above, this representation seems to suggest a *decrease* of the effect of relativity on $|f(r^*)|$ with increasing data resolution after reaching a maximum at approx. $\sin\theta/\lambda=5\text{Å}^{-1}$. One has to note, however, that the formation of this local maximum is due to the convolution of the *increasing* relativistic effects with the systematic *decrease* of $|f(r^*)|$ due to the fact that a Pt atom is not a point-like scatterer for X-rays. This behavior has previously been shown for example by Bučinský *et al.* [20] In order to project out the expected increase of the relativistic effects, one may employ a relative difference $(|f_{DKH2}(r^*)| - |f_{NR}(r^*)|)/|f_{DKH2}(r*)|$ which is depicted in Figure 3b. In this relative representation, a step-wise increase of the relativistic effects on the atomic form factor over the full data resolution range is observed. The insert in Figure 3b depicts the according data up to $\sin\theta/\lambda \leq 4\text{Å}^{-1}$. In this lower resolution range, already three distinct step-wise increases of the relativistic effects on $|f(r^*)|$ can be observed, one additional step then occurring at approx. $10\text{Å}^{-1}$, which is directly followed by a continuous increase starting at about $20\text{Å}^{-1}$ and still continuing at $67\text{Å}^{-1}$. The relative difference approaches a value of 10% at $4\text{Å}^{-1}$ and 80% at the highest data resolution cut-off, respectively, indicating how important the proper treatment of relativistic effects on the inner electronic shells for atomic scattering factors is.

Comparing the radial structure of the atomic scattering factor of the Pt atom in Figure 2 to this representation allows to correlate the five step–wise increases observed for the relativistic effects on $|f(r^*)|$ to the electronic shells of the Platinum atom. For example, the scattering factor contributions of the $n=5$ shell show a severe decay already below $\sin\theta/\lambda=0.5\text{Å}^{-1}$, which directly corresponds to the first increase of the relativistic effects depicted in the insert of Figure 3b at the same data resolution range. A similar relation can be observed for the $n = 4$ shell, for which the scattering factor contributions drop to almost zero just below $1.5\text{Å}^{-1}$, which corresponds to the second sharp increase of the relativistic effects. In this way we may conclude from Figure 3b, that up to a data resolution of $4.0\text{Å}^{-1}$ at most the relativistic effects on three electronic shells $(n = 5, 4, 3)$ in a Platinum atom may be detected by the according step-wise increases in the relativistic effects.

The aim of the present study is to transfer these considerations on atom scattering factors to molecular structure factors and determine the relativistic effects on them with respect to a four–component reference. Moreover, we will comment on the experimental significance of the results obtained. For comparability reasons, we investigated the same molecular model systems studied previously by Eickerling *et al.* with respect to relativistic effects on the topology of the electron density, namely the formally $d^{10}$ metal organic fragments $M(C_2H_2)$ ($M$ = Ni (**1**), Pd (**2**) and Pt (**3**)). [17] These acetylene complexes can for $M$=Ag be stabilized experimentally by bulky ligands [38] and combined experimental and theoretical charge density studies have been performed to investigate the nature of the chemical bonding between the acetylene ligand and the metal atom [38, 39]. For our study of relativistic effects on the structure factors, we employed the quasi–relativistic *Douglas–Kroll–Hess second order (DKH2)* [40–42] and the *zeroth-order regular approximation (ZORA)* [43–45] Hamiltonians in comparison to the "fully relativistic" *Dirac-Coulomb Hamiltonian* (DHF) ‡ and the according non-relativistic limit (denoted as NR in the following).

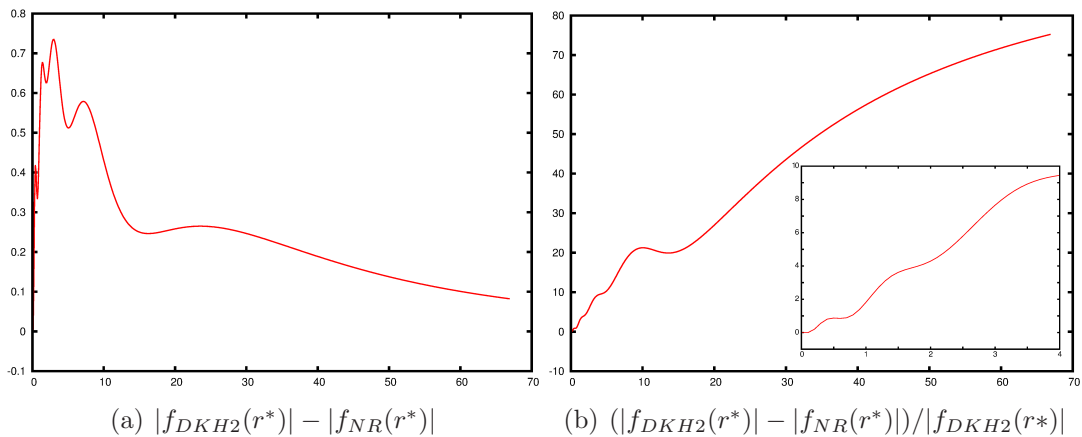

**Figure 3.** Relativistic effects calculated as $|f_{DKH2}(r^*)| - |f_{NR}(r^*)|$ on the atomic form factor of an isolated Platinum atom *vs.* the resolution $\sin\theta/\lambda$ in Å$^{-1}$; (a) absolute values, (b) relative values $(|f_{DKH2}(r^*)| - |f_{NR}(r^*)|)/|f_{DKH2}(r*)|$ given in percent.

## 2. Methods

Molecular Hartree–Fock calculations employing the scalar DKH2, the ZORA (non-scaled, four-component metric), the four–component DHF and the according NR reference Hamiltonians in combination with the fully decontracted quadruple–$\zeta$ basis sets described in Reference [17] have been performed using the DIRAC11 program. [47] The molecular geometries of the compounds **1**–**3** were adapted from Reference [17].

The calculation of static X-ray structure factors $\mathbf{F}_c(\mathbf{r}^*)$ from these molecular wave functions is a straightforward procedure, which first requires the introduction of a pseudo–translational symmetry by defining an arbitrary unit cell for the pseudo–lattice. Once the lattice is defined, numerical [6] or in some cases analytical methods [48–51] can be employed to evaluate the Fourier-transform

$$\mathbf{F}(\mathbf{r}^*) = \int_{cell} \rho(\mathbf{r}) e^{2\pi i \mathbf{r}^* \cdot \mathbf{r}} d^3\mathbf{r} \qquad (3)$$

of the molecular electron density distributions $\rho(\mathbf{r})$. The structure factor $\mathbf{F}(\mathbf{r}^*)$ defined by Equation (3) is a vector in the complex plane of numbers, which reduces to a real number for centrosymmetric crystals. Therefore two molecules have been arranged in the unit cell (orthorhombic, $a = 15$, $b = 5$, $c = 10$Å) to impose inversion symmetry. The total structure factor can be expressed in terms of the structure factor of one of these molecules, for which the Fourier-transform of Equation (3) has to be calculated. [8]

For the purpose of our study a interface between the quantum–chemistry codes and the programs employed for the numerical structure factor calculation was required.

‡ For reviews on the theoretical background of these "four–component" methods, see for example References [46] and [29].

In order to obtain maximum flexibility with respect to the codes providing the wave functions, we implemented a general interface, which for a calculation of static and dynamic structure factors only requires the possibility to calculate $\rho(\mathbf{r})$ at an arbitrary point in space at a time, a feature that is implemented in most of the commonly used quantum–chemistry program packages or the according *quantum theory of atoms in molecules* (QTAIM) [55] routines. Based on this electron density data, we employ the numerical calculation of $\mathbf{F}(\mathbf{r}^*)$ as implemented in the DENPROP code. [6] In this ansatz, $\rho(\mathbf{r})$ is calculated on a grid of points for each pseudo–atom determined by a weighting factor based on the Becke scheme [52] on angular Lebedev [53] and radial Gauß-Chebyshev grids [54].

We note, that the program interface has also been generalized to obtain dynamic structure factors by a folding of the static structure factors with atomic thermal displacement parameters. This requires a proper atomic partitioning scheme for the total molecular electron density similar to the assumption of independent spherical atoms in the scope of the *independent atom model* (IAM). The partitioning of $\rho(\mathbf{r})$ within the QTAIM is not a proper choice in this case, as the convolution of the atomic bassin densities with different Debye-Waller factors would result in discontinuities at the bassin boundaries. We therefore employ the *Stockholder partitioning* [56] of $\rho(\mathbf{r})$ according to

$$\rho_i(\mathbf{r}) = w_i \rho(\mathbf{r}) \quad \text{with } w_i = \frac{\rho_i^0(\mathbf{r})}{\sum_i^N \rho_i^0(\mathbf{r})} \quad (4)$$

where $\sum \rho_i^0(\mathbf{r})$ represents a so called *pro-molecule density*, which results from the superposition of the density distributions $\rho_i^0(\mathbf{r})$ of non–interacting atoms. This partitioning results in *fuzzy* atomic densities $\rho_i(\mathbf{r})$, which are allowed to overlap and therefore can be scaled by different atomic thermal parameters. The Stockholder partitioning was preferred over the Becke scheme we employed for the static structure factor calculations, because of the more flexible wheighting scheme of the former, which explicitly takes the atomic radii into account. This ansatz was validated by employing it for example to the calculated structure factors of the molecular model system $[\text{ScCH}_3]^{2+}$. The **U** parameters resulting from a HC-model refinement (Sc: $U_{11}$=0.028999(1), $U_{22}$=0.034001(1), $U_{33}$=0.023994(1)Å$^2$; C: $U_{11}$=0.043923(2), $U_{22}$=0.052903(2) , $U_{33}$=0.030905(2)Å$^2$; H: $U_{iso}$=0.057553(34)Å$^2$) are in excellent agreement with the according parameters originally employed for the convolution (Sc: $U_{11}$=0.029, $U_{22}$=0.034, $U_{33}$=0.025Å$^2$; C: $U_{11}$=0.042, $U_{22}$=0.053 , $U_{33}$=0.031Å$^2$; H: $U_{iso}$=0.057Å$^2$).§

Atomic form factors have been calculated analytically using DENPROP [6] based on the Hartree–Fock wavefunction provided by Gaussian09 [63] employing a universal Gaussian basis set (UGBS) and the DKH2/NR Hamiltonians. [64]

§ The reference thermal parameters were obtained from experimental data deposited in the CCSD [57] (CCDC RefCodes: EXADAI [58], EXADEM [58], HIMDOX [59], IXURIC [60], PIXGUZ [61], PIXHAG [61], QEYXUO [59], UHIXIS [62], UHIXOY [62]; all data collected at $T = 173$K)

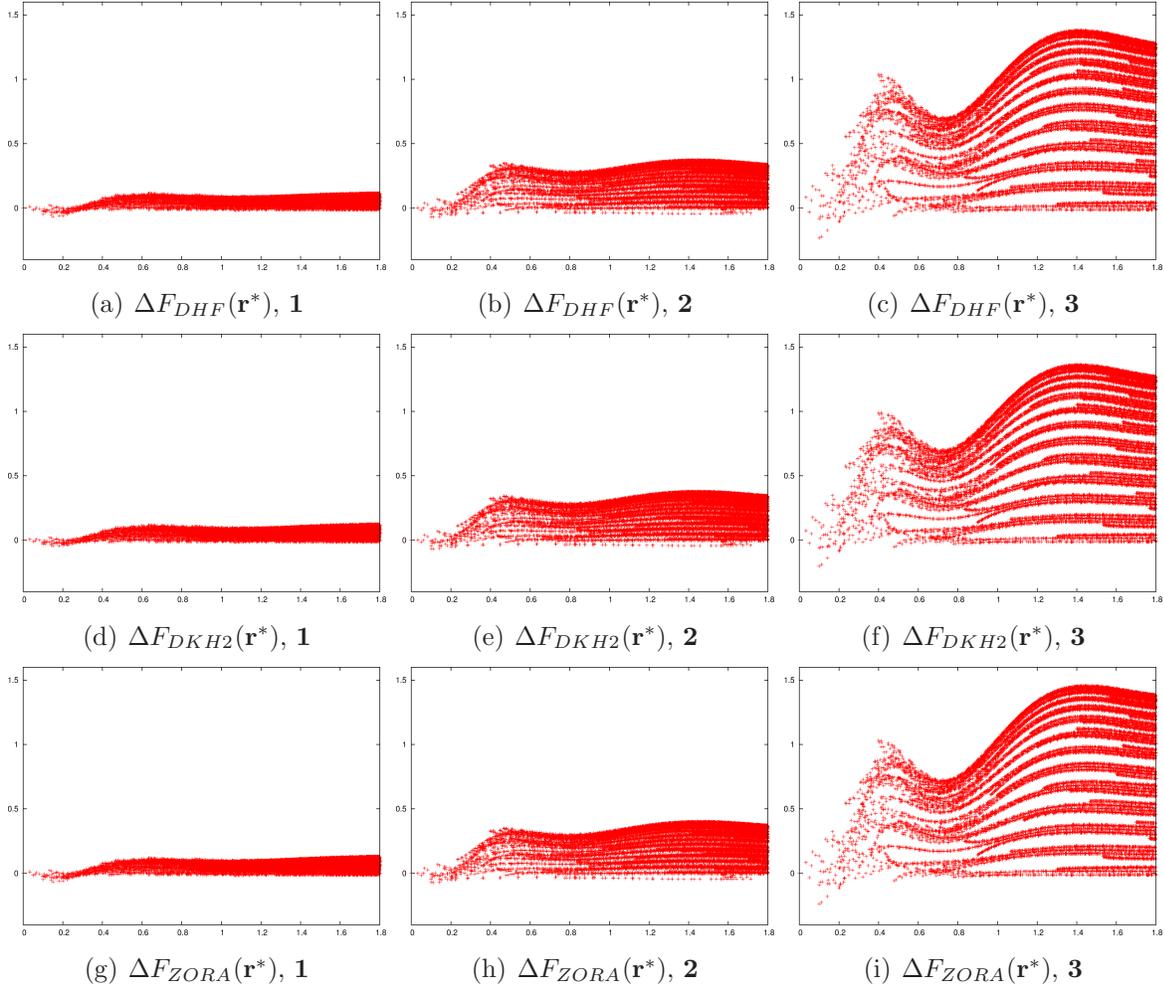

**Figure 4.** Absolute structure factor differences $\Delta F_{DHF}(\mathbf{r}^*)$, $\Delta F_{DKH2}(\mathbf{r}^*)$ and $\Delta F_{ZORA}(\mathbf{r}^*)$ for the model systems **1** (a,d,g), **2** (b,e,h) and **3** (c,f,i) *vs.* the resolution $\sin\theta/\lambda$ in Å$^{-1}$.

## 3. Results and Discussion

Based on the considerations on atomic structure factors presented in Chapter 1, similar results may be expected for the molecular model systems **1**, **2** and **3**, for which the molecular structure factors $\mathbf{F}(\mathbf{r}^*)$ are to a large extend dominated by the atomic scattering contribution of the transition metal atoms. In the following, we will first discuss the relativistic effects on the $\mathbf{F}(\mathbf{r}^*)$ in terms of the absolute differences $\Delta F_{DHF}(\mathbf{r}^*) = |\mathbf{F}_{DHF}(\mathbf{r}^*)| - |\mathbf{F}_{NR}(\mathbf{r}^*)|$, $\Delta F_{DKH2}(\mathbf{r}^*) = |\mathbf{F}_{DKH2}(\mathbf{r}^*)| - |\mathbf{F}_{NR}(\mathbf{r}^*)|$ and $\Delta F_{ZORA}(\mathbf{r}^*) = |\mathbf{F}_{ZORA}(\mathbf{r}^*)| - |\mathbf{F}_{NR}(\mathbf{r}^*)|$ up to a data resolution of $\sin\theta/\lambda < 1.8\text{Å}^{-1}$. The according differences *vs.* the data resolution are depicted in Figure 4.

As it might be expected, the relativistic effects on the structure factors increase with the nuclear charge of the transition metal atom in the series **1-3** (see Figure 4a-c) and the absolute differences reach a maximum of approx. $\Delta F_{DHF}(\mathbf{r}^*) = 1.4$ for **3**. The truncation of the data resolution to the limiting sphere of Ag radiation stresses the limited amount of information available on the inner atomic shells in this case. In all three representations of $\Delta F_{DHF}(\mathbf{r}^*)$, at most two significant increases as signatures of the individual electronic shells of the transition metal atoms (*vide supra*), can be identified. Comparing the differences $\Delta F_{DHF}(\mathbf{r}^*)$ to the according values for $\Delta F_{DKH2}(\mathbf{r}^*)$ (Figure 4d–f) and $\Delta F_{ZORA}(\mathbf{r}^*)$ (Figure 4g–i) indicates that the relativistic effects can be well accounted for by both quasi–relativistic model Hamiltonians. For all three model systems the according scatter–plots show a very similar distribution of points. This may be further exemplified by considering the differences between $|\mathbf{F}_{DHF}(\mathbf{r}^*)| - |\mathbf{F}_{DKH2}(\mathbf{r}^*)|$ and $|\mathbf{F}_{DHF}(\mathbf{r}^*)| - |\mathbf{F}_{ZORA}(\mathbf{r}^*)|$ which are smaller than 0.1 for **3**. These findings are therefore in agreement with the results of a previous study on the relativistic effects on the topology of the electron density in real space, which also showed, that the according relativistic effects are well accounted for by both quasi–relativistic model Hamiltonians [17].

The formation of a local maximum due to the convolution of the overall decay of the structure factor and the increase in the relativistic effects with increasing data resolution discussed before for the Platinum atom is not visible for the molecular structure factors within the given data resolution of $\sin\theta/\lambda < 1.8\text{Å}^{-1}$. Still, in order to allow for assessing the experimental significance of the relativistic effects on the $|\mathbf{F}(\mathbf{r}^*)|$ values later, considering relative differences is more appropriate. Figure 5 therefore depicts the relative differences $\Delta F_{DHF}(\mathbf{r}^*)/|\mathbf{F}_{NR}(\mathbf{r}^*)|$, $\Delta F_{DKH2}(\mathbf{r}^*)/|\mathbf{F}_{NR}(\mathbf{r}^*)|$ and $\Delta F_{ZORA}(\mathbf{r}^*)/|\mathbf{F}_{NR}(\mathbf{r}^*)|$ for the model systems **1**–**3**. Note, that in this representation structure factors $|\mathbf{F}_{NR}(\mathbf{r}^*)|$ smaller than 0.05 have been omitted from the data, because their very small absolute values lead to relative differences of several hundred percent, thus severely biasing the representations shown in Figure 5. These might be considered as statistical outliers, but a closer inspection reveals some very interesting trends for these particular reflections, which we will briefly discuss for the data of model system **3**.

The cutoff eliminates 50 weak reflections ($0.002 < |\mathbf{F}_{DHF}(\mathbf{r}^*)| < 0.05$) from the data set of **3** in a resolution range between 1.0 and 1.6Å$^{-1}$, for which on average $\Delta F_{DHF}(\mathbf{r}^*)/|\mathbf{F}_{NR}(\mathbf{r}^*)| \approx 220\%$. This high average value in turn is mostly due to four individual reflections (the (13 -9 19), (13 9 19), (13 -9 -19) and (13 9 -19)), for which $\Delta F_{DHF}(\mathbf{r}^*)/|\mathbf{F}_{NR}(\mathbf{r}^*)| \approx 2000\%$, while being only 56% for the rest of the 46 weak reflections. Two of these four reflections are symmetry related to each other by the mirror plane in the molecular plane (perpendicular to the unit cell *b*-axis) of **3**, so that the (13 -9 19)/(13 9 19) and (13 -9 -19)/(13 9 -19) have the same values $\mathbf{F}_{DHF}(\mathbf{r}^*)$ of 0.012/0.0136, respectively. For the (13 -9 19), we find $\Delta F_{DHF}(\mathbf{r}^*)/|\mathbf{F}_{NR}(\mathbf{r}^*)| \approx 2600\%$, for the (13 -9 -19) $\Delta F_{DHF}(\mathbf{r}^*)/|\mathbf{F}_{NR}(\mathbf{r}^*)| \approx 1500\%$. We further observe, that the 50 weak reflections and also the 58 reflections with are most affected by relativistic effects (relative differences between 2600% and 21%) all belong to the ($hk|l|$) series (13 *k* 19). The particular *h* and *l* will of course be correlated to the chosen orientation of the molecule in the pseudo-translational unit cell and the according orientation of the lattice plane. However, within this series, severe differences in the values of $\Delta F_{DHF}(\mathbf{r}^*)/|\mathbf{F}_{NR}(\mathbf{r}^*)|$ are observed even for very similar lattice planes, *i.e.* for $k = 7, 8, 9, 10, 11$ we find $\Delta F_{DHF}(\mathbf{r}^*)/|\mathbf{F}_{NR}(\mathbf{r}^*)|$=76, 59, 1600, 111, 58%, respectively. The reason for these changes is not clear up to now, but it might correlate individual reflections to the shell structure of the Pt atom. It is finally interesting to note, that the two reflections (13 -9 19) and (13 -9 -19) in a comparison of a EHC and IAM refinement *vs.* the $F_{DHF}(\mathbf{r}^*)$ are among the ones that yield the largest relative difference $|\mathbf{F}_{EHC}(\mathbf{r}^*)| - |\mathbf{F}_{IAM}(\mathbf{r}^*)|/|\mathbf{F}_{IAM}(\mathbf{r}^*)|$ of all reflections (12%). The maximum difference for this comparison is found for the (13 8 19) reflection with approx. 1278%, a reflection which is again a member of the (13 k 19) series and for which $\Delta F_{DHF}(\mathbf{r}^*)/|\mathbf{F}_{NR}(\mathbf{r}^*)|$ is as high as 60%. These observations might indicate a possible correlation between reflections that are strongly affected by the multipolar expansion of the electron density within the EHC model and those showing pronounced relativistic effects, an aspect which warrants further investigation with respect to the experimental significance of relativistic effects on X-ray structure factors.

Considering the differences $\Delta F_{DHF}(\mathbf{r}^*)/|\mathbf{F}_{NR}(\mathbf{r}^*)|$ of the $|\mathbf{F}_{DHF}(\mathbf{r}^*)| > 0.05$, the $y$–scale of the according scatter plots in Figure 5 indicates, that for all three model systems **1-3** the relativistic effects on the structure factors amount to approx. $\pm 25\%$ maximum. The relative differences therefore do not recover the absolute increase of the relativistic effects between the transition metals in **1**, **2** and **3**. This is obviously due to the increase of $|\mathbf{F}_{NR}(\mathbf{r}^*)|$ along the same row, so that the increase of the structure factor compensates the increase of the relativistic effects. It may therefore be expected, that the increase of the relative relativistic effects in **3** compared to **1** and **2** are only visible at even higher data resolutions.

Comparing these results to the values of $\Delta F_{DKH2}(\mathbf{r}^*)/|\mathbf{F}_{NR}(\mathbf{r}^*)|$ and $\Delta F_{ZORA}(\mathbf{r}^*)/|\mathbf{F}_{NR}(\mathbf{r}^*)|$, the according scatter plots shown in Figure 5d–i confirm again the overall good performance of the quasi–relativistic Hamiltonians. The relative differences obtained for the DKH2 and the ZORA Hamiltonian show a very similar distribution within the according scatter plots, and the relative differences between the DKH2 and the ZORA results to the four-component reference are small and amount to a *maximum* of approx. 2.5% for both, the DKH2 and the ZORA Hamiltonian (*vide infra*).

The differences depicted in Figure 5 may be more quantitatively discussed in terms of crystallographic $R$-values. Computing the $R_1$ values for $N$ reflections according to

$$R_1 = \frac{\sum_N ||F_{DHF/DKH2/ZORA}(\mathbf{r}^*)| - |F_{NR}(\mathbf{r}^*)||}{\sum_N |F_{NR}(\mathbf{r}^*)|} \quad (5)$$
$$= \frac{\sum_N |\Delta F_{DHF/DKH2/ZORA}(\mathbf{r}^*)|}{\sum_N |F_{NR}(\mathbf{r}^*)|}$$

results in $R_1$=0.81, 1.51 and 2.78% for the difference $\Delta F_{DHF}(\mathbf{r}^*)$ for **1**, **2** and **3**, respectively. Also with respect to these quality criteria, the relativistic effects are well described by the quasi–relativistic Hamiltonians, since the according $R_1$=0.84, 1.53 and

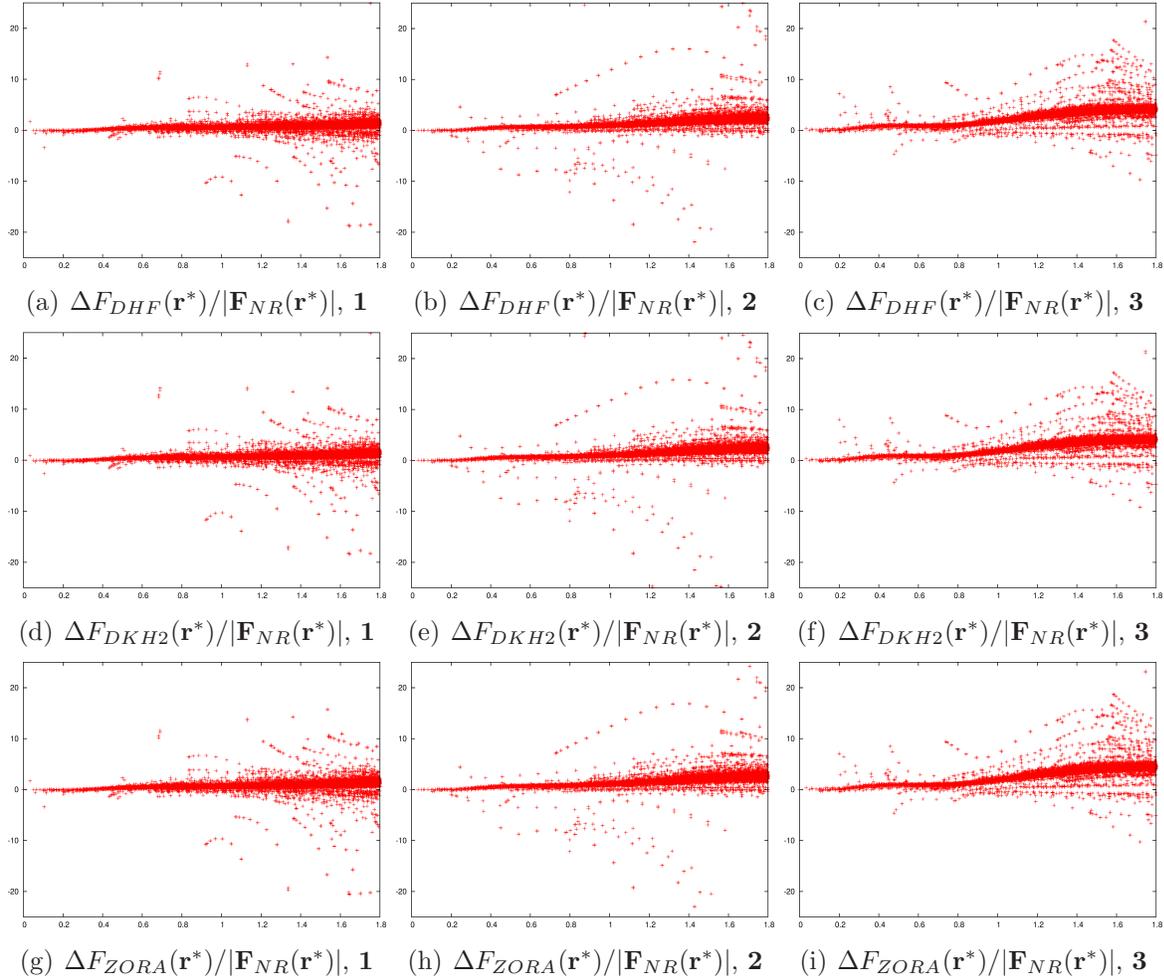

**Figure 5.** Relative structure factor differences $\Delta F_{DHF}(\mathbf{r}^*)/|\mathbf{F}_{NR}(\mathbf{r}^*)|$, $\Delta F_{DKH2}(\mathbf{r}^*)/|\mathbf{F}_{NR}(\mathbf{r}^*)|$ and $\Delta F_{ZORA}(\mathbf{r}^*)/|\mathbf{F}_{NR}(\mathbf{r}^*)|$ for the model systems **1** (a,d,g), **2** (b,e,h) and **3** (c,f,i) *vs.* the resolution $\sin\theta/\lambda$ in Å$^{-1}$. Structure factors with $|\mathbf{F}_{NR}(\mathbf{r}^*)| < 0.05$ have been omitted for clarity (see text).

2.76% and $R_1$=0.90, 1.62 and 2.94% for $\Delta F_{DKH2}(\mathbf{r}^*)$ and $\Delta F_{ZORA}(\mathbf{r}^*)$, respectively, are very similar to the $R_1$-values of $\Delta F_{DHF}(\mathbf{r}^*)$.

Summarizing the results obtained so far we note, that the relativistic effects on X-ray structure factors of model compounds **1** to **3** can amount to approx. 20–25%. For individual reflections (or reflection series) we find that relativistic effects can even be orders of magnitude larger, and the possible relation between these pronounced relativistic effects and the importance of these reflections for the multipolar expansion of the electron density within the HC model warrants further investigation of these observations with respect to their experimental significance. Employing a quasi–relativistic Hamiltonian takes most of these effects into account, so that the maximum differences comparing the DKH2 or ZORA results to the four–component reference amount to approx. 2.5% for reflections with $|\mathbf{F}_{DHF}(\mathbf{r}^*)| > 0.05$. This is also true for the most affected reflections with $\Delta F_{DHF}(\mathbf{r}^*)/|\mathbf{F}_{NR}(\mathbf{r}^*)|$ for **3** being as high as 2600%, for which a value of $\Delta F_{DKH2}(\mathbf{r}^*)/|\mathbf{F}_{NR}(\mathbf{r}^*)|$=2578% for example indicates the very good performance of the DKH2 model Hamiltonian in recovering the relativistic effects.

These results should be comparable to the previous study of the relativistic effects on the topology of the electron density on the same model compounds. [17, 65]‖ In this study, the relativistic effects on the electron density of the $M$–C bond critical points $\rho(\mathbf{r}_{BCP})$ were found to be 0, 1.3 and 6% for **1**, **2** and **3**, respectively, which is indeed comparable to the 0.81, 1.51 and 2.78% mentioned above. However, opposite to the values of $\rho(\mathbf{r}_{BCP})$, the structure factor data presented above not only contains local information on the valence but rather on the total electron density distributions of the molecules. We therefore prefer a direct comparison of the according difference electron density maps as presented in Reference [17] and [65] to the according difference Fourier maps obtained from the structure factor differences. We find, however, that such a direct comparison is to some extend not possible. Calculating the according difference Fourier maps from the structure factors obtained from the DHF and the NR Hamiltonians (Figure 6a-c) leads to difference maps, which are contaminated by severe Fourier truncation artifacts. This is most obvious for the according difference Fourier maps obtained from $\mathbf{F}_{DHF}(\mathbf{r}^*) - \mathbf{F}_{NR}(\mathbf{r}^*)$, which are so much distorted by Fourier artifacts, that for **2** and **3** not even the positions of the carbon and hydrogen atoms can be identified without problems (see Figure 6a-c). These artifacts are usually only obtained when a direct Fourier transform of a (limited) set of $\mathbf{F}(\mathbf{r}^*)$ is calculated, while for a *difference Fourier map*, the according ripples should cancel each other. The reason for this not being the case for the difference Fourier maps presented in Figure 6a–c lies in the relativistic effects discussed above. Due to the significant difference in the radial distribution of the NR and the (quasi-)relativistic electron densities, the Fourier ripples do obviously no longer cancel each other. Therefore, in order to obtain meaningful Fourier maps, one should not consider the comparison of the (quasi-)relativistic to the NR result, but rather the comparison of (for example) the DHF with the DKH2 results. Note, however, that these maps no longer depict *relativistic effects*, but rather illustrate the extend, to which the quasi–relativistic Hamiltonian reproduces the four–component result.

This conclusion (and the overall good performance of the DKH2 Hamiltonian) is supported by the observation, that the difference maps employing $\mathbf{F}_{DHF}(\mathbf{r}^*)-\mathbf{F}_{DKH2}(\mathbf{r}^*)$ are indeed not affected by severe Fourier artifacts. In contrast, these maps (see Figure 6d–f) reveal interesting non–radially symmetric features which may indeed be compared to the according difference density maps directly calculated from the according wave functions presented in Figure 6 g–i. These features hint at a significant influence of relativistic effects on the valence shell charge concentrations [66–68] of the transition metal atoms, which are not fully recovered by the DKH2 Hamiltonian. [17] We note in passing, that also effects due to the limited data resolution are clearly visible by the truncation of the local maxima close to the Ni atom position in Figure 6d compared to Figure 6g. In addition, the according map depicted in Figure 6f reveals significant differences between the four–component reference and the approximate DKH2 Hamiltonian, even in the carbon–metal bonding region of **3**. However, the non–zero contours visible in the $M$–C bonding region in Figure 6f are not reproduced in the according difference density map (Figure 6i). This indicates, that the relativistic effects in the $M$–C bonding region of **3** which are described in [17] are to a large amount recovered by the DKH2–Hamiltonian and that the non-zero contours in this region (Figure 6f) are caused by remaining Fourier artifacts, which are of much smaller magnitude than those occurring due to the relativistic effects.

‖ Taking the overall similar performance of the ZORA and DKH2 Hamiltonian into account, we will in the following exemplarily employ the DKH2 results for the further discussion.

Analyzing in more detail the differences between the structure factors obtained from the DKH2 and the DHF Hamiltonian, one interesting aspect is the overall larger relative difference $(|\mathbf{F}_{DHF}(\mathbf{r}^*)|-|\mathbf{F}_{DKH2}(\mathbf{r}^*)|)/|\mathbf{F}_{DHF}(\mathbf{r}^*)|$ for **1** compared to **2** and **3** (see Figure 7). Quantitatively, the according $R_1$-values are found to be 0.04, 0.02 and 0.03%, for **1**, **2** and **3**, respectively. This observation may be rationalized by taking the data resolution and the radial extend of the according transition metal atoms into account. As the inner electronic shells of a Nickel atom which are most affected by relativistic effects are less compact in direct space, their contribution to the X-ray structure factor is larger in the resolution range considered than for Palladium and Platinum. Again we note, that this is only true for the medium data resolutions considered within the present study and different trends must be expected for higher data resolutions.

In the following, we will finally discuss the magnitude of the relativistic effects on $|\mathbf{F}(\mathbf{r}^*)|$ with respect to their experimental significance. In particular, we will focus on the comparison of relativistic effects on $\mathbf{F}(\mathbf{r}^*)$ to the according effects due to chemical bonding as determined by *experimental* charge density studies. For this purpose, we analyzed a previously published experimental X-ray structure factor data set from the literature, namely of the complex $[Ag(C_2H_2)(Al(OC(CF_3)_3)_3)_4]$ **4** [38,39]. This complex contains the same $M(C_2H_2)$ fragment as our model systems **1**–**3** and the according data set has been collected for the purpose of an experimental charge density study, therefore providing the required data quality, redundancy and completeness (up to a resolution of 1.1 Å$^{-1}$). In order to determine, whether relativistic effects in the order of magnitude of 1-3% are of any experimental significance, we may evaluate the effects of a multipolar modelling of the experimental data within a HC model on the values of $|\mathbf{F}_c(\mathbf{r}^*)|$ compared to an *independent atom model* (IAM). The according *relative* differences $(|\mathbf{F}_{HC}(\mathbf{r}^*)| - |\mathbf{F}_{IAM}(\mathbf{r}^*)|)/|\mathbf{F}_{IAM}(\mathbf{r}^*)|$ for the data set of **4** are depicted in Figure 8. As one can clearly see, the changes in the calculated structure factors taking charge transfer and aspherical density distributions into account amounts on average to only a few percent. In particular, 19967 of the 22357 reflections (89%) show a relative difference smaller than 2.5% (Figure 8b). The latter value might thus be employed as a measure for the data accuracy which is routinely achievable nowadays by X-ray diffraction experiments. From this consideration we might conclude, that the relativistic effects on X-ray structure factors as discussed above are of the same order of magnitude as charge transfer and chemical bonding effects on the electron density. We furthermore may compare the $R_1$ values introduced above to quantify the differences also for this experimental dataset. Employing the HC-model structure factors as reference to compare to the IAM data, we obtain a value of $R_1 = 0.86\%$ for the experimental dataset of **4**. Limiting the data resolution of the calculated data to $1.1$Å$^{-1}$, we obtain $R_1 = 0.47$, 0.80 and 1.27% for **1**, **2** and **3**, respectively for the differences $\Delta F_{DHF}(\mathbf{r}^*)$. The $R_1$ values for the relativistic effects in **2** (0.80%) and the effects of the multipolar expansion of the structure factors for **4** (0.86%) are not only of the same order of magnitude but indeed almost identical. This comparison therefore supports our conclusion, that relativistic effects should clearly be extractable from experimental structure factor data providing the required data accuracy for the experimental determination of electron density distributions.

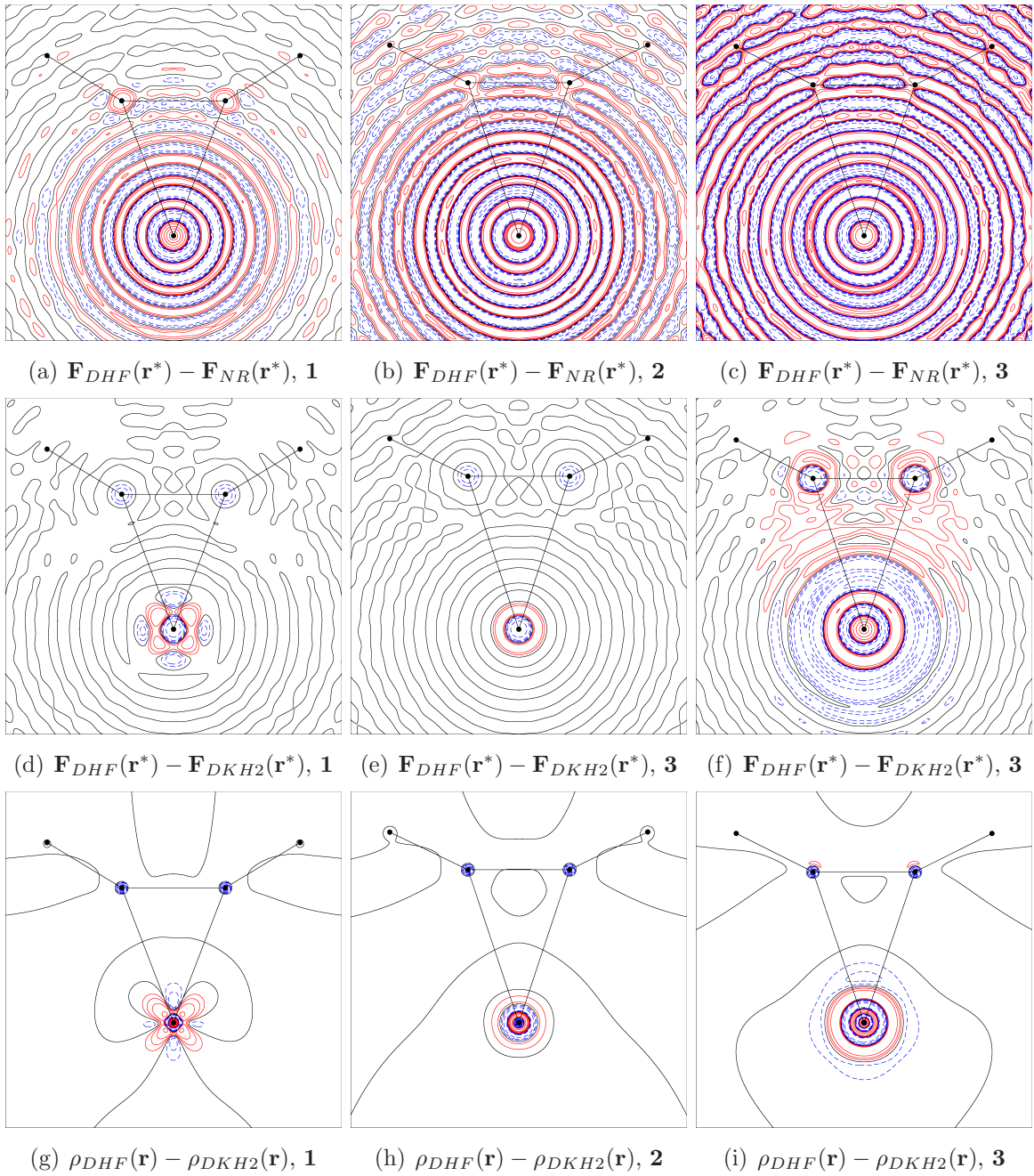

**Figure 6.** Fourier-Transform to real space of the (phased) difference of the structure factors $\mathbf{F}_{DHF}(\mathbf{r}^*) - \mathbf{F}_{NR}(\mathbf{r}^*)$ and $\mathbf{F}_{DHF}(\mathbf{r}^*) - \mathbf{F}_{DKH2}(\mathbf{r}^*)$ for compound **1** (a,d), **2** (b,e) and **3** (c,f); panels g–h depict the difference densities $\rho_{DHF}(\mathbf{r}) - \rho_{DKH2}(\mathbf{r})$ in the molecular plane of **1**–**3**. Atomic positions are marked by a filled circle, contour values at $\pm 0, 2, 4, 8 \cdot 10^n$ ($n = -2, -1, 0, 1, 2, 3$)$e\text{Å}^{-3}$, positive and negative values are drawn as red solid and blue dashed lines, respectively; zero contour as black solid line.

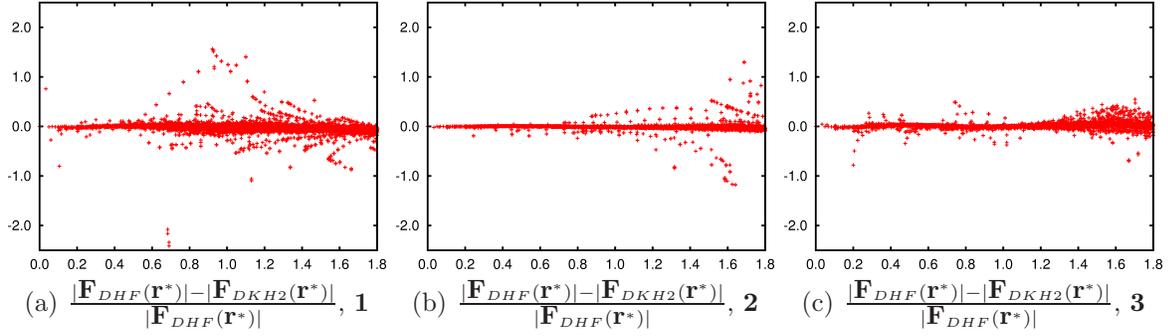

**Figure 7.** Relative structure factor differences $(|\mathbf{F}_{DHF}(\mathbf{r}^*)| - |\mathbf{F}_{DKH2}(\mathbf{r}^*)|)/|\mathbf{F}_{DHF}(\mathbf{r}^*)|$ for the model systems **1** (a), **2** (b) and **3** (c) *vs.* the resolution $\sin\theta/\lambda$ in Å$^{-1}$. In order to remove outliers, structure factors with $|\mathbf{F}_{DHF}(\mathbf{r}^*)| < 0.05$ have been omitted.

We finally note, that the according $R_1$-values for the difference $|\mathbf{F}_{DKH2}(\mathbf{r}^*)| - |\mathbf{F}_{NR}(\mathbf{r}^*)|$ (0.48, 0.81, 1.26% for **1**, **2** and **3**, respectively) are almost identical and indicate, that the differences between different (quasi–)relativistic model Hamiltonians will only play a minor role for experimental studies. However, the *maximum* differences for some reflections reach values of 2% (see Figure 7a–c) and 22% (*vide supra*) which may be indeed of importance when aiming at the reconstruction of $\rho(\mathbf{r})$ at *subatomic resolution*, *i.e.* when explicitly aiming at effects like contractions or polarizations of the inner electronic shells. [7, 8, 65] This is illustrated by the fact that especially for **2** the maximal deviation in Figure 7b occurs for data resolutions $\sin\theta/\lambda > 1.4$Å$^{-1}$, *i.e.* in a data range required especially for charge density studies at subatomic resolution. [8] This result warrants further investigation on relativistic effects with respect to the atomic wave function data employed for the multipolar expansion of the electron density for such studies.

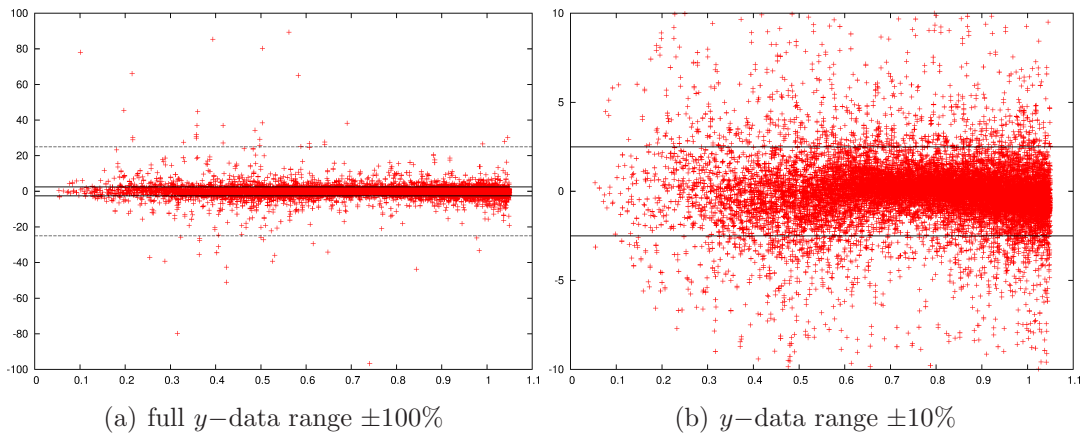

(a) full $y$–data range $\pm 100\%$

(b) $y$–data range $\pm 10\%$

**Figure 8.** Relative structure factor difference $(|\mathbf{F}_{HC}(\mathbf{r}^*)| - |\mathbf{F}_{IAM}(\mathbf{r}^*)|)/|\mathbf{F}_{IAM}(\mathbf{r}^*)|$ for the experimental structure factors of **4** *v.s.* the data resolution $\sin\theta/\lambda$ in Å$^{-1}$; (a) full data range; (b) zoom to $\pm 10\%$.

## 4. Summary

In summary, we presented in this work results of a systematic study on relativistic effects on calculated molecular X-ray scattering factors. The comparison of the structure factors obtained from the non–relativistic and the four–component calculations demonstrates, that for the model systems **1**–**3** these effects amount to 0.81, 1.51 and 2.78% in the resolution range of $\sin\theta/\lambda < 1.8$Å$^{-1}$, respectively. For individual reflections or reflection series, the effects can be as high as several hundred percent, a fact that warrants further investigation with respect to the significance of such reflections for the reconstruction of electron density distributions from the structure factor data via a HC model. Comparing the according results of the three model systems **1-3** it was shown, that due to the different radii of the transition metal atoms Ni, Pd and Pt the *relative* magnitudes of the relativistic effects are rather similar for first-, second- and third-row transition metal compounds in the resolution range studied. Employing quasi–relativistic Hamiltonians such as the DKH2– and the ZORA–Hamiltonian leads to a reduction of the differences relative to the four–component results by one order of magnitude. This is in line with the findings of the previous study on the relativistic effects in real space, which have also shown, that for **1**–**3** relativistic effects can be well described by these model Hamiltonians. [17]

We further demonstrated, that a comparison of the difference density maps obtained from the (quasi-)relativistic structure factor data to the ones obtained from the real space electron density distributions [17] is to some extend impossible. Due to the scalar relativistic contraction of the inner electronic shells of the transition metal atoms in **1**–**3**, a difference Fourier transformation of the structure factors obtained by applying different model Hamiltonians only provides reasonable data when comparing the quasi–relativistic to the four–component results. The relativistic effects, *i.e.* the changes of the radial extend of the electronic shells when comparing the NR to the four–component electron density distributions are so pronounced, that the according difference maps are severely affected by Fourier artifacts.

We finally investigated the experimental significance of the above results. By considering the structure factor differences which occur for experimental data of **4** when comparing a HC–model to an IAM ($R_1 = 0.86\%$) we could demonstrate, that the magnitude of the relativistic effects for the calculated data of **2** is indeed almost identical ($R_1 = 0.80\%$) and thus lies well within the data accuracy that is required and routinely achievable today for an experimental charge density study. As we could further demonstrate, the maximum differences between the structure factors obtained from the DKH2 and the DHF Hamiltonian can be as large as 2.5–22% for some reflections. This might provide first evidence, that four–component wave function databases should be employed for the (E)HC-modeling of structure factors at very high resolutions if the aim of the study is for example the determination of inner shell polarization and contraction effects in transition metal compounds.